\documentclass[twocolumn,prl,showpacs,preprintnumbers,amsmath,amssymb]{revtex4}
\usepackage{graphicx}% Include figure files \usepackage{dcolumn}
\usepackage{mathptmx,bm}
\begin{document}

%\preprint{APS/123-QED}

\title{Ultrafast Melting of Spin Density Wave Order in BaFe$_{2}$As$_{2}$ Observed by Time- and Angle-Resolved Photoemission Spectroscopy with Extreme-Ultraviolet Higher Harmonic Generation}

\author{H. Suzuki$^{1}$,  K. Okazaki$^{2}$, T. Yamamoto$^{2}$, T. Someya$^{2}$, M. Okada$^{2}$, K. Koshiishi$^{1}$,  M. Fujisawa$^{2}$, T. Kanai$^{2}$, N. Ishii$^{2}$, M. Nakajima$^{3}$, H. Eisaki$^{3}$, K. Ono$^4$, H. Kumigashira$^4$, J. Itatani$^{2}$, A. Fujimori$^{1}$ and S. Shin$^{2}$}
 
\affiliation{$^{1}$Department of Physics, University of Tokyo,
Bunkyo-ku, Tokyo 113-0033, Japan}

\affiliation{$^{2}$Institute for Solid State Physics (ISSP), University of Tokyo, Kashiwa, Chiba 277-8581, Japan}

\affiliation{$^{3}$National Institute of Advanced Industrial Science and Technology (AIST), Tsukuba, Ibaraki 305-8568, Japan}

\affiliation{$^4$KEK, Photon Factory, Tsukuba, Ibaraki 305-0801, Japan}

\date{\today}
% It is always \today, today,
%  but any date may be explicitly specified

\begin{abstract}
Transient single-particle spectral function of BaFe$_{2}$As$_{2}$, a parent compound of iron-based superconductors, has been studied by time- and angle-resolved photoemission spectroscopy with an extreme-ultraviolet laser generated by higher harmonics from Ar gas, which enables us to investigate the dynamics in the entire Brillouin zone.
We observed electronic modifications from the spin-density-wave (SDW) ordered state within $\sim$ 1 ps after 
the arrival of a 1.5 eV pump pulse. We observed optically excited electrons at the zone center above $E_{F}$  at 0.12 ps, and their rapid decay. After the fast decay of the optically excited electrons, a thermalized state appears and survives for a relatively long time. From the comparison with the density-functional theory band structure for the paramagnetic and SDW states,  we interpret the experimental observations as the melting of the SDW. Exponential decay constants for the thermalized state to recover back to the SDW ground state are $\sim$ 0.60 ps both around the zone center and the zone corner. 

\end{abstract}

 %Pacs changed. Nov 19 2015
% PACS, the Physics and Astronomy
% Classification Scheme. %\keywords{Suggested keywords}%Use show keys class option if keyword
%display desired

\maketitle

%Introduction of pump-probe experiments

Ultrafast phenomena in condensed matter have been the subject of intense research. In order to investigate nonequilibrium transient states in solids, the pulsed photons are divided into pump and probe portions: the pump pulses excite the system and the probe pulses measure various physical quantities of the transient states after variable delay times. On the other hand, angle-resolved photoemission spectroscopy (ARPES) is a versatile tool to study the electronic structure of solids with momentum resolution. Combining these strengths, time- and angle-resolved photoemission spectroscopy (TrARPES) was  realized shortly after the advent of laser-based ARPES. TrARPES technique has been utilized to study ultrafast dynamics of various materials including charge-density-wave materials \cite{Schmitt.F_etal.Science2008,Petersen.J_etal.Phys.-Rev.-Lett.2011}, cuprate superconductors \cite{Graf.J_etal.Nat-Phys2011}, graphene \cite{Johannsen.J_etal.Phys.-Rev.-Lett.2013,Ulstrup.S_etal.Phys.-Rev.-Lett.2014}, and topological insulators \cite{Sobota.J_etal.Phys.-Rev.-Lett.2013,Yamamoto.T_etal.Phys.-Rev.-B2015}.

%introduction of ba122 and its 
%BaFe$_{2}$As$_{2}$ (Ba122), one of the parent compounds of iron-based superconductors (FeSCs), exhibits stripe-type antiferromagnetic (AFM) order below the N\'eel temperature ($T_{N}$) of 138 K. Nesting between the hole Fermi surfaces (FSs) at the zone center and electron ones at the zone corner is considered to enhance the spin-density-wave instability with the ordering vector ${\bm Q}=(\pi/a,\pi/a,2\pi/c)$. The AFM order coexists with the orthorhombic lattice distortions, which entails the orbital differentiation between the $d_{xz}$ and $d_{yz}$ orbitals. Due to the availability of high-quality single crystals, the equilibrium electronic and magnetic structures of Ba122 and those of the doped supercondutors have been extensively studied in order to clarify the competition and interplay between several order parameters \cite{Canfield.P_etal.Annu.-Rev.-Condens.-Matter-Phys.2010,Lu.D_etal.Annual-Review-of-Condensed-Matter-Physics2012,Dai.P_etal.Rev.-Mod.-Phys.2015}. 

%Apparatus
%Since its discovery, 
The ultrafast dynamics of iron-based superconductors (FeSCs) has also been the focus of intense research. In particular, coherent $A_{1g}$ phonon oscillations of the As atoms have been observed by various pump-probe experiments. Mansart \textit{et al.} \cite{Mansart.B_etal.Phys.-Rev.-B2009} reported oscillatory components overlaid on exponential decay in the transient reflectivity spectra of Ba(Fe$_{0.92}$Co$_{0.08}$)$_{2}$As$_{2}$. TrARPES studies on Ba/EuFe$_{2}$As$_{2}$ \cite{Rettig.L_etal.Phys.-Rev.-Lett.2012,Avigo.I_etal.J.-Phys.-Condens.-Matter2013} found oscillations of the chemical potential of the electrons after the pump pulse. Time-resolved x-ray diffraction experiments \cite{Rettig.L_etal.Phys.-Rev.-Lett.2015,Gerber.S_etal.Nat-Commun2015} measured the oscillations in the intensity of the Bragg reflections from BaFe$_{2}$As$_{2}$ (Ba122). %Although the existence of coherent lattice dynamics has been well established so far, a few questions remain. Since the initial process is the absorption of photons by the excitation of valence electrons, electron distribution should be observed well above $E_{F}$, up to the photon energy ($h\nu$) of pump beam. Furthermore, since it is not to the crystal lattice but to the electronic system that the optical pump directly transfers its energy, modifications in the electronic structure should precede the subsequent excitation of phonons, as described by phenomenological two-temperature models \cite{Allen.P_etal.Phys.-Rev.-Lett.1987,Perfetti.L_etal.Phys.-Rev.-Lett.2007}. 
A natural question is how the electronic system is modified concomitantly with the lattice. Since the temperature scale of the pump beam (1.5 eV $\sim$ 17400 K) is much higher than $T_{N}$, it may drive the electronic system to undergo a phase transition to the paramagnetic (PM) state. In equilibrium, the reconstruction of the band structure across the SDW transition has been clearly observed by ARPES by changing temperature \cite{Yi.M_etal.Phys.-Rev.-B2009}. Therefore, one expects that the phase transition can be induced by optical pumping and the relaxation to the SDW ground state is detectable by TrARPES. Furthermore, since investigations into SDW materials by TrARPES have been limited, TrARPES studies on Ba122 will give us insight into the dynamics of SDW formation in general. %Furthermore, since the relaxation dynamics is strongly influenced by the coupling between valence electrons and low-energy excitations like spin waves and lattice vibrations, the information about the nonequilibrium states may give us insight into the pairing mechanism of high-$T_{c}$ superconductivity. 

%In order to perform TrARPES, probe photons with $h\nu$ larger than the work functions of solids are required, which is conventionally achieved by wavelength conversion using nonlinear crystals like $\beta$-BaB$_{2}$O$_{4}$; however, the available photon energies are typically below 6-7 eV, limiting accessible momentum space to the region near the $\Gamma$ point. 
Conventional TrARPES typically utilizes 6-7 eV laser achieved by wavelength conversion.
For  the FeSCs, however, the information about electron pockets around the X point cannot be obtained. To overcome this hurdle, we have employed higher harmonic generation (HHG) by focusing fundamental light to a cell filled with rare gas \cite{Shen.Y_etal.2003}. This method can create photons in the extreme ultraviolet regime ($h\nu>20$ eV), which are high enough to probe the entire Brillouin zone (BZ) of FeSCs. A previous work by Yang \textit{et al.} \cite{Yang.X_etal.Phys.-Rev.-Lett.2014} utilized a similar setup and observed a global oscillation of the chemical potential due to the $A_{1g}$ phonon mode of As atoms both at the $\Gamma$ and X points. However, the time-resolved data in that work were limited to the evolution of angle-integrated photoemission intensity.

\begin{figure}[htbp] %  figure placement: here, top, bottom, or page
   \centering
   \includegraphics[width=8cm]{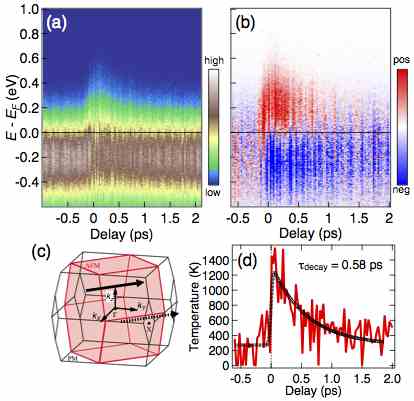} 
   \caption{(Color online) (a) Temporal evolution of angle-integrated photoemission intensity around the $(0,0,6.5)$ point of the Brillouin zone (BZ). (b) Difference spectra of (a) from the average intensity before the arrival of pump pulse. (c) First BZs of the paramagnetic (black) and antiferromagnetic (red) states. We shall use the paramagnetic notation below. Solid and dotted arrows indicate the momentum cuts for the $(0,0,6.5)$ and $(1,1,6.1)$ points, respectively. (d) Temporal evolution of the electronic temperature ($T_{e}$). $T_{e}$ is evaluated by fitting the photoemission intensity to the Fermi-Dirac distribution function convoluted with the instrumental gaussian function. Black dotted line shows the fit to a decay function of the form $T_{e}(t)=A+B\Theta(t)\exp(-t/\tau_{\text{decay}})$, where $\Theta(t)$ is the Heaviside step function.} 
   \label{PESGamma}
\end{figure}

In the present work, we have performed TrARPES measurements of Ba122 in the SDW state in order to directly measure transient single-particle spectral functions both around the zone center and zone corner. The fundamental laser system is a 1 kHz Ti:Sapphire laser operating at wavelengths ($\lambda$) of 800 nm with a pulse width of 90 fs. The second harmonics with $\lambda=$ 400 nm generated by a 0.5-mm-thick $\beta$-BaB$_{2}$O$_{4}$ crystal are employed as the source of HHG. Higher harmonics (HHs) were generated by irradiating the second harmonics to a cell filled with Ar gas and the 9th harmonics with $h\nu =$ 28 eV were selected by a pair of SiC/Mg multilayer mirrors by suppressing other HHs \cite{Ishizaka.K_etal.Phys.-Rev.-B2011}. %Ba122 single crystals were cleaved \textit{in situ} prior to measurements to obtain fresh surfaces. Cleavage occurs along the $ab$ plane. 
Ba122 Single crystals were not detwinned and, therefore, the ARPES intensity is the superposition of intensities from two inequivalent domains in the SDW state. The kinetic energies and the momenta of photoelectrons were measured using a Scienta R4000 hemispherical electron energy analyzer. The total energy resolution was 250 meV. All the measurements were done at 10 K, which was well below the $T_{N}$ of the equilibrium SDW transition. %The structure of Ba122 in the PM state has a space group symmetry of I4/mmm and its
The first BZs for the PM and antiferromagnetic (AFM) states are shown in Fig. \ref{PESGamma} (c). For simplicity, we shall use the notation of the PM BZ to specify positions in the momentum space. In-plane ($k_{X}$, $k_{Y}$) and out-of-plane  ($k_{z}$) momenta are expressed in units of $\pi/a$ and $2\pi/c$, respectively, where $a=3.97$ \AA \,and $c=13.00$ \AA\, are the in-plane and out-of-plane lattice constants, respectively. %The $k_{X}$, $k_{Y}$, and $k_{z}$ axes are also shown in Fig. \ref{PESGamma} (c). 

%Result Gamma point

Figure \ref{PESGamma} (a) shows the temporal evolution of angle-integrated photoemission (PES) spectra around the zone center. With $h\nu = 28 $ eV probe photons the momentum cut is on the $k_{z}=6.5 (2\pi/c)$ plane as shown by a solid arrow in Fig. \ref{PESGamma} (c). One observes that upon the arrival of a pump pulse at $t=0$, the electron population above the Fermi level ($E_{F}$) suddenly increases and it decays afterwards. To improve the visibility of the data, we show the difference spectra in Fig. \ref{PESGamma} (b). Here, the spectrum average before $t=0$ has been subtracted. It is clear that just after $t=0$ the population above $E_{F}$ increases while that below $E_{F}$ decreases, and that this population modification decreases with delay time. In order to quantify the transient state, we show in Fig. \ref{PESGamma} (d) the electronic temperature $T_{e}$ as a function of delay time. $T_{e}$ was determined by fitting the PES intensity at each delay time to a Fermi-Dirac distribution function convoluted with  a Gaussian with the instrumental resolution of 250 meV. After the pump pulse, $T_{e}$ reaches $\sim 1400$ K. $T_{e}$ was fitted to a decay function of the form $T_{e}(t)=A+B\Theta(t)\exp(-t/\tau_{\text{decay}})$, where $\Theta(t)$ is the Heaviside step function. The best fit is shown by a dotted curve. We obtained a decay constant of $\tau_{\text{decay}}=0.58$ ps around the (0,0,6.5) point.
 
\begin{figure*}[htbp] %  figure placement: here, top, bottom, or page
   \centering
   \includegraphics[width=16cm]{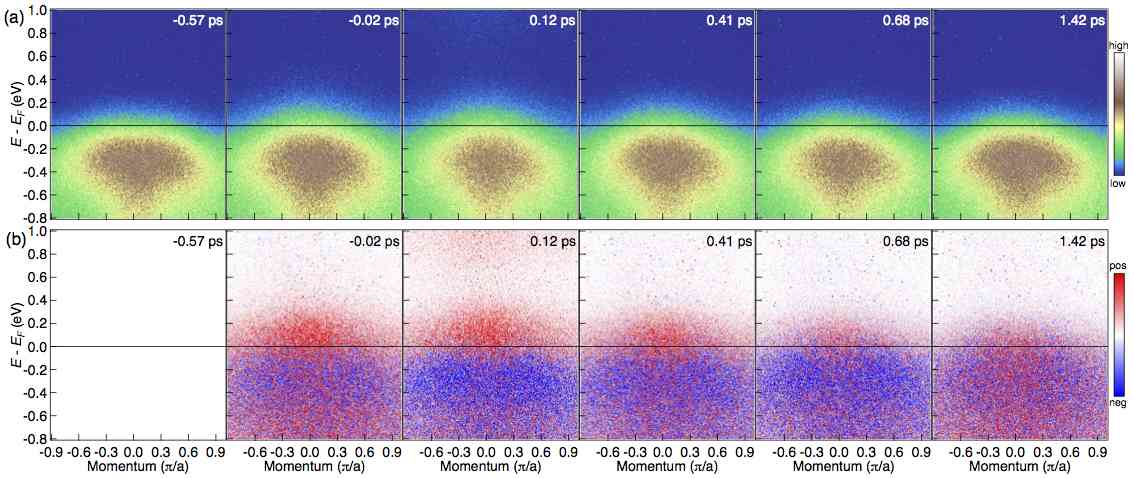} 
   \caption{(Color online) (a) Time-resolved angle-resolved photoemission (TrARPES) spectra around the $(0,0,6.5)$ point. (b) Difference TrARPES spectra from the spectrum before the pump arrival ($t=-0.57$ ps).}
   \label{TrARPESG}
\end{figure*}

In order to study the momentum-resolved spectral function during the relaxation, we show in Fig. \ref{TrARPESG} (a) the TrARPES spectra taken at different delay times. The momentum cut is the same as that of Fig. \ref{PESGamma}. To highlight the change of the spectra, we also show the difference TrARPES spectra in Fig. \ref{TrARPESG} (b). Here, the spectrum at -0.47 ps has been subtracted from each spectrum.
Before the pump arrival (-0.47 ps), the spectrum should represent that of the SDW state, although the identification is not obvious under the present experimental resolution. We later argue that the spectrum represents that of the SDW state. After the arrival of pump pulse ($t\geq$ $-0.02$ ps), electrons start to occupy states above $E_{F}$. At 0.12 ps, high-energy states are populated above the gap ($E-E_{F}>0.6$ eV) and the states have parabolic, electron-band-like intensity distribution. At 0.41 ps, this population disappears, indicating the fast decay of electrons in the high energy state. As for the intensity between $0<E-E_{F}<0.4$ eV, it already exists at -0.02 ps, reaches maximum around 0.12 ps, and slowly decays as a function of time.
At 1.42 ps, there is almost no separation of positive and negative area in the difference spectrum, indicating that the electronic system has relaxed back to the equilibrium state.

\begin{figure}[htbp] %  figure placement: here, top, bottom, or page
   \centering
   \includegraphics[width=8cm]{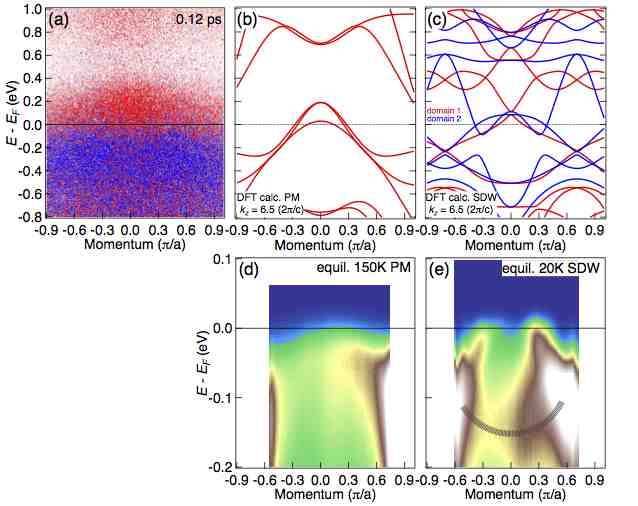} 
   \caption{(Color online) (a) Difference TrARPES spectrum at 0.12 ps. The blue dotted curve is a guide to the eye indicating a characteristic parabolic intensity decrease.  (b), (c) Density-functional theory (DFT) band structure around the (0,0,6.5) for paramagnetic (PM) [(b)] and spin-density-wave (SDW) [(c)] states. For the SDW state, band structures for two inequivalent domains are plotted. (d), (e) Equilibrium ARPES spectra around the $Z$ point for the PM (200 K) and SDW (20 K) states, respectively. Black dotted line in panel (e) indicates a parabolic feature which appears only in the SDW state.}
   \label{compeq}
\end{figure}

Here we discuss the implications of the present data for an ultrafast phase transition. 
In the following, we argue that the observed hot electron states are those of the PM states caused by the melting of the SDW state. We compare in Figs. \ref{compeq} (a) and (b) the difference spectrum at 0.12 ps with density-functional theory (DFT) band structure at $k_{z}=6.5$ ($2\pi/c$) in the PM state calculated within the generalized gradient approximation (GGA) \cite{Perdew.J_etal.Phys.-Rev.-Lett.1996}. The intensity increase is shown by red in the difference spectrum in Fig. \ref{compeq} (a). We can confirm high-energy population above 0.6 eV and a gap between $0.4<E-E_{F}<0.6$ eV. This is qualitatively consistent with the calculated PM band structure shown in Fig. \ref{compeq} (b), and quantitatively the smaller gap in experiment can be ascribed either to the renormalization of the band structure or to the experimental resolution. We also show the DFT band structures for the SDW state in Fig. \ref{compeq} (c). Since the Ba122 samples were not detwinned, we show the results for two inequivalent domains. %It is well known that electronic structure calculations of the parent compounds of FeSCs greatly overestimate the AFM ordered moment \cite{Yin.Z_etal.Phys.-Rev.-Lett.2008,Mazin.I_etal.Nat.-Phys.2009}, and the artificial reduction of the ordered moment by the LDA+$U$ method with negative $U$ leads to a large change of the band structure \cite{Yi.M_etal.Phys.-Rev.-B2009}. However, Terashima \textit{et al.} \cite{Terashima.T_etal.Phys.-Rev.-Lett.2011} showed that the standard DFT calculation reasonably accounts for the AFM FSs of Ba122 determined by the Shubnikov-de Haas oscillation measurements. Therefore we have used the standard GGA functional and reproduced the results of Terashima \textit{et al.}. 
One observes that the band structure near $E_{F}$ in the SDW state is more entangled than that in the PM state, due to the band folding caused by the AFM ordering. In particular, the band bottoms of domain 1 at 0.3 eV and the flat band at 0.6 eV of domain 2 are not clearly resolved in the present measurement.  From these comparisons, the electronic states at 0.12 ps can be identified as that of the PM state, as expected from the high $T_{e}>1000$ K after the optical pump.

To gain further insight from the intensity decrease shown in blue, equilibrium spectra taken with $h\nu=63$ eV (Z point) for the SDW (20K) and PM (150 K) states are shown in Figs. \ref{compeq} (d) and (e), respectively.
We note that the blue intensity in panel (a) shows a parabolic, electron-band-like feature. We ascribe this to a parabolic feature observed in the equilibrium the SDW state as indicated by a black dotted line in Fig. \ref{compeq} (e), which does not appear in the PM state [Fig. \ref{compeq} (d)]. This feature originates from the electron band bottom located at the zone corner in the PM state. The disappearance of this feature by raising the temperature from the SDW state to the PM state is also reported in Ref. \onlinecite{Yi.M_etal.Phys.-Rev.-B2009}. Thus the data at -0.47 ps can be naturally assigned to those of the SDW state, as expected from the sample temperature of 10 K $< T_{N}$. 

%The observation of high energy states at the conduction band bottom is possible due to the relatively slow relaxation of excited electrons across the energy gap at the zone center. Indeed, hot carriers after optical pump was also observed in bilayer graphene but not in monolayer graphene \cite{Johannsen.J_etal.Phys.-Rev.-Lett.2013,Ulstrup.S_etal.Phys.-Rev.-Lett.2014}. Here, the existence of this band gap in bilayer graphene leads to longer-lifetime carriers at the minimum of the conduction band than those in the Dirac cone dispersions in the monolayer graphene. 

\begin{figure}[htbp] %  figure placement: here, top, bottom, or page
   \centering
   \includegraphics[width=8cm]{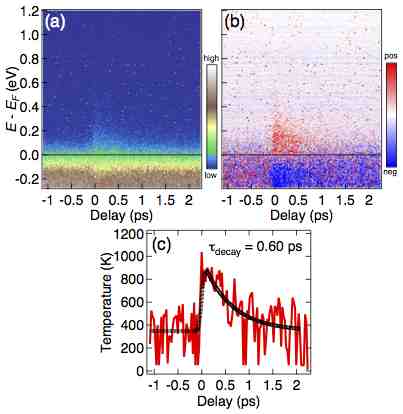} 
   \caption{(Color online) (a) Temporal evolution of angle-integrated PES spectra around the $(1,1,6.1)$ point. The momentum cut is indicated by a dotted arrow in Fig. \ref{PESGamma} (c). (b) Difference spectra of (a) from the average intensity before the arrival of pump pulse. (c) Temporal evolution of $T_{e}$.}
   \label{decayele}
\end{figure}

%X point
Next we investigate the dynamics of excited electrons around the zone corner. Figure \ref{decayele} (a) shows the temporal evolution of angle-integrated PES spectra at the (1,1,6.1) point. We employed $h\nu=28$ eV and the momentum cut in the BZ is indicated by a dotted arrow in Fig. \ref{PESGamma} (c). The difference spectra are also shown in Fig. \ref{decayele} (b). As in the case of the (0,0,6.5) point, the intensity above $E_{F}$ increases and that below $E_{F}$ decreases at $t=0$, and this change weakens with time. The evolution of $T_{e}$ is plotted in Fig. \ref{decayele} (c). $T_{e}$ rises up to $\sim$ 900 K just after the pump pulse. We obtain a decay constant of $0.60$ ps, close to $0.58$ ps around the (0,0,6.5) point.

\begin{figure*}[htbp] %  figure placement: here, top, bottom, or page
   \centering
   \includegraphics[width=16cm]{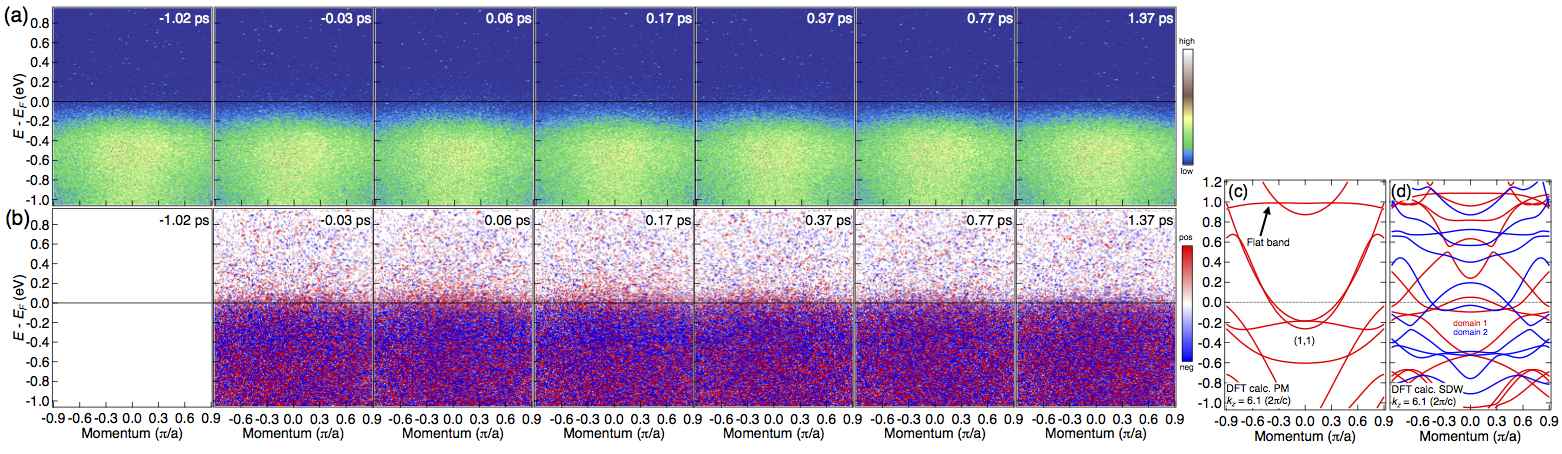} 
   \caption{(Color online) (a) TrARPES spectra around the $(1,1,6.1)$ point. (b) Difference TrARPES spectra from the spectrum before the pump arrival ($t=-1.02$ ps). (c), (d) DFT band structures around the $(1,1,6.1)$ point for the PM and SDW states, respectively.}
   \label{TrARPESX}
\end{figure*}

TrARPES spectra around the (1,1,6.1) point are shown in Fig. \ref{TrARPESX} (a) and the difference spectra are shown in Fig. \ref{TrARPESX} (b). Although the modifications in the spectra are weaker than those in the zone center, we observed an intensity increase after the pump pulse. Note that we did not observe high-energy population well above $E_{F}$ within our measurement range. To understand the difference between the results for the (0,0,6.5) and (1,1,6.1) points, we show the calculated PM band structure in Fig. \ref{TrARPESX} (c). %The lowest states above $E_{F}$ at (1,1,6.1) point are located at 0.873 eV above $E_{F}$, which is much higher than those at the $(0,0,6.5)$ point. Furthermore, 
One of the two bands located around $\sim$ 1 eV above $E_{F}$ is holelike and flat; as a result, the lifetime of excited electrons around the (1,1,6.1) point can be much shorter than around that at (0,0,6.5), because they can easily decay into different momentum states via electron-phonon and/or electron-electron interactions. To support the interpretation, we also show the calculated SDW band structure for two domains in  Fig. \ref{TrARPESX} (d). As in the (0,0,6.5) point, there are more bands in the SDW state than in the PM state. In particular, there are three electron band bottoms between $0.2<E-E_{F}<0.6$ eV, where long excitation lifetime is expected. The absence of the observation of transient population in this energy range corroborates our assignment of the transient states between 0.1 and 0.2 ps to the PM state. %Furthermore, we note that our two momentum cuts shown in Fig. \ref{PESGamma} (c) would be nearly equivalent in the SDW state. If the SDW state had survived the pump pulse, we would have observed the population inversion as around the (0,0,6.5) point. The absence of this observation, therefore, corroborates our assignment of the transient state between 0.1 and 0.2 ps to the PM state.

While the high-energy electrons above 0.6 eV around $(0,0,6.5)$ point decay quickly ($<0.41$ ps), the hot electrons close to $E_{F}$ persist for a longer time, which can be interpreted as the thermalized electronic states. For the time scale of the relaxation of near-$E_{F}$ states, we compare the present decay constants with those from previous works. The present value $\sim$ 0.60 ps for 10 K is close to 0.8 ps for EuFe$_{2}$As$_{2}$ \cite{Rettig.L_etal.Phys.-Rev.-Lett.2012}, 0.38 ps for Ba122 \cite{Yang.X_etal.Phys.-Rev.-Lett.2014} . Also, it is comparable with 0.6-0.8 ps for superconducting Ba$_{1-x}$K$_{x}$Fe$_{2}$As$_{2}$ \cite{Chia.E_etal.Phys.-Rev.-Lett.2010} estimated from time-resolved reflectivity. As for the A$_{1g}$ phonon mode, we did not observe any oscillatory intensity modulation near $E_{F}$. Mansart \textit{et al.} \cite{Mansart.B_etal.Phys.-Rev.-B2009} %performed time-resolved reflectivity measurements of Ba(Fe$_{0.92}$Co$_{0.08}$)$_{2}$As$_{2}$ with various pump fluences and 
found that, while the phonon oscillation is barely visible at $\sim 1.3$ mJ/cm$^{2}$ in the course of the exponential decay of the reflectivity, it becomes more prominent at higher flux ($\sim 2$ mJ/cm$^{2}$) and makes the relaxation slower. The absence of oscillations may suggest that our data represents the electronic state in weakly pumped regime. Other possibilities include the polarization and the width of the pump pulse. We need further investigation into the dependence of the transient states on fluence, polarization and pump width in the future.%In the presence of the present setup, we could not achieve higher flux due to sudden increase of multiphoton photoemission, which distorts the spectra by space-charge effects.
%conclusion

In conclusion, we have studied the ultrafast dynamics of the BaFe$_{2}$As$_{2}$ in the SDW state by TrARPES using a HH from rare gas. %The probe photon energy of $h\nu = 28$ eV created by HHG from Ar gas is high enough to investigate the dynamics in the entire Brillouin zone. 
We observed electronic modifications from the SDW band structure within $\sim$ 1 ps after 
the 1.5 eV pump pulse. High-energy states above $E_{F}$ were observed at 0.12 ps at the zone center and they decay rapidly. The 0.12 ps spectrum is more consistent with the band structure calculation of the PM state than that of the SDW state, which is expected from the high $T_{e}$ above 1000 K. After the fast decay of the optically excited electrons, a thermalized state appears and survives for a relatively long time. Decay constants both around the $\Gamma$ point and the X point are $\sim$ 0.60 ps, in agreement with previous TrARPES and time-resolved reflectivity measurements.

%\section{ACKNOWLEDGEMENTS} 

This work was supported by JSPS KAKENHI (Grant No. JP26610095) and Photon and Quantum Basic Research Coordinated Development Program from the Ministry of Education, Culture, Sports,
Science and Technology, Japan. Experiment at Photon Factory was approved by the Photon Factory Program Advisory Committee (Proposal No. 2014G177). H.S. acknowledges financial support from Advanced Leading Graduate Course for Photon Science (ALPS) and the JSPS Research Fellowship for Young Scientists. 

\bibliography{TrPESBa122}

\end{document}